\documentclass[preprint,showkeys,aps]{revtex4}
\usepackage{amssymb}
\usepackage{amsmath}
\usepackage{amsfonts}
\usepackage{graphicx}

\newtheorem{theorem}{Theorem}

\newtheorem{definition}[theorem]{Definition}

\begin{document}

\title{Scaling property and the generalized entropy uniquely\\
determined by a fundamental nonlinear differential equation}
\author{Hiroki Suyari}
\email{suyari@faculty.chiba-u.jp, suyari@ieee.org}
\affiliation{Department of Information and Image Sciences, Chiba University, Chiba
263-8522, Japan}
\author{Tatsuaki Wada}
\email{wada@ee.ibaraki.ac.jp}
\affiliation{Department of Electrical and Electronic Engineering, Ibaraki University,
Hitachi, Ibaraki 316-8511, Japan}
\pacs{PACS number}

\begin{abstract}
We derive a scaling property from a fundamental nonlinear differential
equation whose solution is the so-called $q$-exponential function. A scaling
property has been believed to be given by a power function only, but
actually more general expression for the scaling property is found to be a
solution of the above fundamental nonlinear differential equation. In fact,
any power function is obtained by restricting the domain of the $q$%
-exponential function appropriately. As similarly as the correspondence
between the exponential function and Shannon entropy, an appropriate
generalization of Shannon entropy is expected for the scaling property.
Although the $q$-exponential function is often appeared in the optimal
distributions of some one-parameter generalized entropies such as R\'{e}nyi
entropy, only Tsallis entropy is uniquely derived from the algebra of the $q$%
-exponential function, whose uniqueness is shown in the two ways in this
paper.
\end{abstract}

\date{\today }
\maketitle

\section{Scaling property derived from a fundamental nonlinear differential
equation}

The exponential function is often appeared in every scientific field. Among
many properties of the exponential function, the linear differential
function $dy/dx=y$ is the most important characterization of the exponential
function. A slightly nonlinear generalization of this linear differential
equation is given by%
\begin{equation}
\frac{dy}{dx}=y^{q}\quad \left( q\in \mathbb{R}\right) .
\label{nonlinear differential equation}
\end{equation}%
(See the equation (17) at page 5 of \cite{GT04} and\ the equations (22)-(23)
at page 8 of \cite{Ts04PD}.) This nonlinear differential equation is
equivalent to%
\begin{equation}
\int \frac{1}{y^{q}}dy=\int dx.  \label{diffequ1}
\end{equation}%
Then we define the so-called $q$\textit{-logarithm} $\ln _{q}x.$%
\begin{equation}
\ln _{q}x:=\frac{x^{1-q}-1}{1-q}
\end{equation}%
as a generalization of $\ln x$. Applying the property:

\begin{equation}
\frac{d}{dx}\ln _{q}x=\frac{1}{x^{q}},
\end{equation}%
to (\ref{diffequ1}), we obtain%
\begin{equation}
\ln _{q}y=x+C  \label{diffequ2}
\end{equation}%
where $C$ is \textit{any} constant \cite{SW06}. Then we define the so-called 
$q$\textit{-exponential} $\exp _{q}\left( x\right) $ as the inverse function
of $\ln _{q}x$ as follows:%
\begin{equation}
\exp _{q}\left( x\right) :=\left\{ 
\begin{array}{ll}
\left[ 1+\left( 1-q\right) x\right] ^{\frac{1}{1-q}} & \text{if }1+\left(
1-q\right) x>0, \\ 
0 & \text{otherwise.}%
\end{array}%
\right.  \label{q-exponential}
\end{equation}%
Note that the $q$-logarithm and $q$-exponential recover the usual logarithm
and exponential when $q\rightarrow 1$, respectively (See the pages 84-87 of 
\cite{AO01} for the detail properties of these generalized functions $\ln
_{q}x$ and $\exp _{q}\left( x\right) $). Thus, the general solution to the
nonlinear differential equation (\ref{nonlinear differential equation})
becomes%
\begin{equation}
y=\exp _{q}\left( x+C\right) =\exp _{q}\left( C\right) \exp _{q}\left( \frac{%
x}{\left( \exp _{q}\left( C\right) \right) ^{1-q}}\right)
\end{equation}%
where $C$ is \textit{any} constant satisfying $1+\left( 1-q\right) C>0.$
Dividing the both sides by $\exp _{q}\left( C\right) $ of the above
solution, we obtain 
\begin{equation}
\frac{y}{\exp _{q}\left( C\right) }=\exp _{q}\left( \frac{x}{\left( \exp
_{q}\left( C\right) \right) ^{1-q}}\right) .  \label{scaling-1}
\end{equation}%
Under the following scaling:%
\begin{equation}
y\prime :=\frac{y}{\exp _{q}\left( C\right) },\quad x\prime :=\frac{x}{%
\left( \exp _{q}\left( C\right) \right) ^{1-q}},  \label{scaling}
\end{equation}%
we obtain%
\begin{equation}
y\prime =\exp _{q}\left( x\prime \right) .
\end{equation}%
This means that the solution of the nonlinear differential equation (\ref%
{nonlinear differential equation}) obtained above is \textit{%
\textquotedblleft scale-invariant\textquotedblright } under the above
scaling (\ref{scaling}). Moreover, we can choose \textit{any} constant $C$
satisfying $1+\left( 1-q\right) C>0$ because $C$ is an integration constant
of (\ref{diffequ1}).

Note that the above scaling (\ref{scaling}) with respect to both variables $%
x $ and $y$ can be observed only when $q\neq 1$ and $C\neq 0$. In fact, when 
$q=1,$ i.e., $y=\exp \left( x+C\right) ,$ (\ref{scaling}) reduces to the
scaling with respect to only $y,$ i.e., $x\prime =x$, and when $C=0,$ both
scalings in (\ref{scaling}) disappears \cite{SW06}.

The above scaling property of the nonlinear differential equation (\ref%
{nonlinear differential equation}) is very significant in the fundamental
formulations for every generalization based on (\ref{nonlinear differential
equation}). We summarize the important points in the above fundamental
result.

\begin{enumerate}
\item $C$ in the scaling (\ref{scaling}) is \textit{any} constant satisfying 
$1+\left( 1-q\right) C>0$ because $C$ is an integration constant of (\ref%
{diffequ1}). This means that the scaling (\ref{scaling}) is \textit{arbitrary%
} for \textit{any} $q$ and $C$ if $q$ and $C$ satisfies $1+\left( 1-q\right)
C>0$ $\left( q\neq 1\text{ and }C\neq 0\right) $.

\item In general studies of differential equation, $C$ is determined by the 
\textit{initial condition} of the nonlinear differential equation (\ref%
{nonlinear differential equation}). This means, when an observable in a
dynamics grows according to the nonlinear differential equation (\ref%
{nonlinear differential equation}), \textit{the initial condition determines
the scaling of the dynamics.} This is applicable to the analysis of the
chaotic dynamics \cite{TPZ97}\cite{LBRT00}\cite{BR02a}\cite{BR02b}\cite{Ro04}%
\cite{BR04}\cite{TT06}.

\item In general, for a mapping $f:X\rightarrow Y$, $f:X\prime \left(
\subset X\right) \rightarrow Y\,$\ is called a restriction of a mapping $f$
to $X\prime $, which is denoted by $f\upharpoonright X\prime $. Let $f_{q}$
be a $q$-exponential function (\ref{q-exponential}) from $\mathbb{R}%
\rightarrow \mathbb{R}^{+}$. For the restricted domain $\mathbb{R\prime }%
_{q} $ defined by%
\begin{equation}
\mathbb{R\prime }_{q}:=\left\{ x\in \mathbb{R\,}\left\vert \mathbb{\,}\left(
1-q\right) x\gg 1\right. \right\} \left( \subset \mathbb{R}\right) ,
\end{equation}%
a restriction of a mapping $f_{q}$ to $\mathbb{R\prime }_{q}$ is denoted by%
\begin{equation}
f\prime _{q}:=f_{q}\upharpoonright \mathbb{R\prime }_{q}:\mathbb{R\prime }%
_{q}\rightarrow \mathbb{R}^{+}.
\end{equation}%
Then $f\prime _{q}$ becomes a power function:%
\begin{equation}
f\prime _{q}\left( x\right) =\left( \left( 1-q\right) x\right) ^{\frac{1}{1-q%
}}\approx x^{\frac{1}{1-q}}.  \label{f'_q}
\end{equation}%
In the above formulations the only case $q<1$ is discussed . As shown in the
section IV, the $q$-generalizations along the line of (\ref{nonlinear
differential equation}) has a symmetry $q\leftrightarrow 2-q$, i.e., $%
1-\left( 1-q\right) \leftrightarrow 1+\left( 1-q\right) $. Therefore, the
above $f\prime _{q}\left( x\right) $ can be replaced by a restriction of a
mapping $f_{2-q}$ to $\mathbb{R\prime }_{2-q}:$ 
\begin{equation}
f\prime _{2-q}\left( x\right) =\left( \left( q-1\right) x\right) ^{\frac{1}{%
q-1}}\approx x^{\frac{1}{q-1}}
\end{equation}%
in accordance with the symmetry, which implies that the case $q>1$ can be
discussed. In this way, the \textit{restriction} of the $q$-exponential
function to the domain $\mathbb{R\prime }_{q}\left( q<1\right) $ or $\mathbb{%
R\prime }_{2-q}\left( q>1\right) $ coincide with a power function, which has
been often appeared and discussed in science. In general, the restricted
domain $\mathbb{R\prime }_{q}$ or $\mathbb{R\prime }_{2-q}$ is called
\textquotedblleft \textit{scaling domain}\textquotedblright\ and its
corresponding range is called \textquotedblleft \textit{scaling range}%
\textquotedblright .

\item A power function $f:\mathbb{R}\rightarrow \mathbb{R}$ is known to be
characterized by the following functional equation, i.e., there exists a
function $g:\mathbb{R}\rightarrow \mathbb{R}$ such that 
\begin{equation}
f\left( bx\right) =g\left( b\right) f\left( x\right)
\end{equation}%
holds for any $b,x\in \mathbb{R}$. The above functional equation uniquely
determines a power function 
\begin{equation}
f\left( x\right) =f\left( 1\right) x^{-\alpha }  \label{power-func}
\end{equation}%
for choosing $g\left( b\right) =b^{-\alpha }$. See some references such as 
\cite{NW05} for the proof. On the other hand, a $q$-exponential function is
characterized by the nonlinear differential equation (\ref{nonlinear
differential equation}) as similarly as a exponential function. Moreover,
the solutions of (\ref{nonlinear differential equation}) are scale-invariant
under the scaling (\ref{scaling}) and reduce to power functions when the
domain is restricted to $\mathbb{R\prime }_{q}$ or $\mathbb{R\prime }_{2-q}$
as shown above. In fact, by restricting the domain of (\ref{scaling-1}) to $%
\mathbb{R\prime }_{q},$ the general solution (\ref{scaling-1}) of the
nonlinear differential equation (\ref{nonlinear differential equation})
reduces to the following power function according to (\ref{f'_q}).%
\begin{equation}
\frac{y}{\exp _{q}\left( C\right) }=\left( \left( 1-q\right) \frac{x}{\left(
\exp _{q}\left( C\right) \right) ^{1-q}}\right) ^{\frac{1}{1-q}}=\frac{%
\left( 1-q\right) ^{\frac{1}{1-q}}}{\exp _{q}\left( C\right) }x^{\frac{1}{1-q%
}},
\end{equation}%
that is,%
\begin{equation}
y=\left( 1-q\right) ^{\frac{1}{1-q}}x^{\frac{1}{1-q}},  \label{reduc-q-exp}
\end{equation}%
which is equivalent to (\ref{power-func}) if 
\begin{equation}
\alpha =\frac{1}{q-1}.
\end{equation}%
Therefore, many discussions on \textquotedblleft exponential versus
power-law\textquotedblright , i.e., \textquotedblleft $\frac{dy}{dx}=y$
versus $f\left( bx\right) =g\left( b\right) f\left( x\right) $%
\textquotedblright\ should be replaced by \textquotedblleft exponential
versus $q$-exponential\textquotedblright , i.e., \textquotedblleft $\frac{dy%
}{dx}=y$ versus $\frac{dy}{dx}=y^{q}$\textquotedblright ,\ which is more
natural\ from mathematical point of view.
\end{enumerate}

As shown in these discussions, the fundamental nonlinear differential
equation (\ref{nonlinear differential equation}) provides us with not only
the characterization of the $q$-exponential function but also the scaling
property in its solution.

As similarly as the relation between the exponential function $\exp \left(
x\right) $ and Shannon entropy, we expect the corresponding information
measure to the $q$-exponential function $\exp _{q}\left( x\right) $. There
exist some candidates such as R\'{e}nyi entropy, Tsallis entropy and so on.
But the algebra derived from the $q$-exponential function uniquely
determines Tsallis entropy as the corresponding information measure. In the
following sections of this paper, we present the two mathematical results to
uniquely determine Tsallis entropy by means of the already established
formulations such as the $q$-exponential law, the $q$-multinomial
coefficient and $q$-Stirling's formula.

\section{$q$-exponential law}

The exponential law plays an important role in mathematics, so this law is
also expected to be generalized based on the $q$-exponential function $\exp
_{q}\left( x\right) $. For this purpose, the new multiplication operation $%
\otimes _{q}$ is introduced in \cite{NMW03} and \cite{Bo03} for satisfying
the following identities:%
\begin{align}
\ln _{q}\left( x\otimes _{q}y\right) & =\ln _{q}x+\ln _{q}y, \\
\exp _{q}\left( x\right) \otimes _{q}\exp _{q}\left( y\right) & =\exp
_{q}\left( x+y\right) .
\end{align}%
The concrete form of the $q$-logarithm or $q$-exponential has been already
given in the previous section, so that the above requirements as $q$%
-exponential law leads us to the definition of $\otimes _{q}$ between two
positive numbers.

\begin{definition}
For two positive numbers $x$ and $y$, the $q$-\textit{product} $\otimes _{q}$
is defined by%
\begin{equation}
x\otimes _{q}y:=\left\{ 
\begin{array}{ll}
\left[ x^{1-q}+y^{1-q}-1\right] ^{\frac{1}{1-q}}, & \text{if }%
x>0,\,y>0,\,x^{1-q}+y^{1-q}-1>0, \\ 
0, & \text{otherwise.}%
\end{array}%
\right.  \label{def of q-product}
\end{equation}
\end{definition}

The $q$-\textit{product} recovers the usual product such that $\underset{%
q\rightarrow 1}{\lim }\left( x\otimes _{q}y\right) =xy$. The fundamental
properties of the $q$-product $\otimes _{q}$ are almost the same as the
usual product, but 
\begin{equation}
a\left( x\otimes _{q}y\right) \neq ax\otimes _{q}y\quad \left( a,x,y\in 
\mathbb{R}\right) .
\end{equation}%
The other properties of the $q$-\textit{product} are available in \cite%
{NMW03} and \cite{Bo03}.

In order to see one of the validities of the $q$-product, we recall the well
known expression of the exponential function $\exp \left( x\right) $ given by%
\begin{equation}
\exp \left( x\right) =\underset{n\rightarrow \infty }{\lim }\left( 1+\frac{x%
}{n}\right) ^{n}.  \label{def of expx}
\end{equation}%
Replacing the power on the right side of (\ref{def of expx}) by the $n$
times of the $q$-product $\otimes _{q}^{n}:$%
\begin{equation}
x^{\otimes _{q}^{n}}:=\underset{n\text{ times}}{\underbrace{x\otimes
_{q}\cdots \otimes _{q}x}},
\end{equation}%
$\exp _{q}\left( x\right) $ is obtained. In other words, $\underset{%
n\rightarrow \infty }{\lim }\left( 1+\frac{x}{n}\right) ^{\otimes _{q}^{n}}$
coincides with $\exp _{q}\left( x\right) .$ 
\begin{equation}
\exp _{q}\left( x\right) =\underset{n\rightarrow \infty }{\lim }\left( 1+%
\frac{x}{n}\right) ^{\otimes _{q}^{n}}  \label{repre of q-exp}
\end{equation}%
The proof of (\ref{repre of q-exp}) is given in the appendix of \cite{Su04b}%
. This coincidence (\ref{repre of q-exp}) indicates a validity of the $q$%
-product. In fact, the present results in the following sections reinforce
it.

\section{$q$-multinomial coefficient and $q$-Stirling's formula}

We briefly review the $q$-multinomial coefficient and the $q$-Stirling's
formula by means of the $q$-product $\otimes _{q}$. As similarly as for the $%
q$-product, $q$\textit{-ratio} is introduced as follows:

\begin{definition}
For two positive numbers $x$ and $y$, the inverse operation to the $q$%
-product is defined by 
\begin{equation}
x\oslash _{q}y:=\left\{ 
\begin{array}{ll}
\left[ x^{1-q}-y^{1-q}+1\right] ^{\frac{1}{1-q}}, & \text{if }%
x>0,\,y>0,\,x^{1-q}-y^{1-q}+1>0, \\ 
0, & \text{otherwise}%
\end{array}%
\right.
\end{equation}%
which is called $q$\textit{-ratio} in \cite{Bo03}.
\end{definition}

$\oslash_{q}$ is also derived from the following satisfactions, similarly as
for $\otimes_{q}$ \cite{NMW03}\cite{Bo03}. 
\begin{align}
\ln_{q}\left( x\oslash_{q}y\right) & =\ln_{q}x-\ln_{q}y, \\
\exp_{q}\left( x\right) \oslash_{q}\exp_{q}\left( y\right) & =\exp
_{q}\left( x-y\right) .
\end{align}

The $q$-product and $q$-ratio are applied to the definition of the $q$%
-multinomial coefficient \cite{Su04b}.

\begin{definition}
For $n=\sum_{i=1}^{k}n_{i}$ and $n_{i}\in \mathbb{N\,}\left( i=1,\cdots
,k\right) ,$ the $q$-multinomial coefficient is defined by%
\begin{equation}
\left[ 
\begin{array}{ccc}
& n &  \\ 
n_{1} & \cdots & n_{k}%
\end{array}%
\right] _{q}:=\left( n!_{q}\right) \oslash _{q}\left[ \left(
n_{1}!_{q}\right) \otimes _{q}\cdots \otimes _{q}\left( n_{k}!_{q}\right) %
\right] .  \label{def of q-multinomial coefficient}
\end{equation}
\end{definition}

From the definition (\ref{def of q-multinomial coefficient}), it is clear
that%
\begin{equation}
\underset{q\rightarrow 1}{\lim }\left[ 
\begin{array}{ccc}
& n &  \\ 
n_{1} & \cdots & n_{k}%
\end{array}%
\right] _{q}=\left[ 
\begin{array}{ccc}
& n &  \\ 
n_{1} & \cdots & n_{k}%
\end{array}%
\right] =\frac{n!}{n_{1}!\cdots n_{k}!}.
\end{equation}

In addition to the $q$-multinomial coefficient, the $q$-Stirling's formula
is useful for many applications such as our main results. By means of the $q$%
-product (\ref{def of q-product}), the $q$-factorial $n!_{q}\,$\ is
naturally defined as follows.

\begin{definition}
For a natural number $n\in \mathbb{N}$ and $q\in \mathbb{R}^{+}$, the $q$%
-factorial $n!_{q}$ is defined by%
\begin{equation}
n!_{q}:=1\otimes _{q}\cdots \otimes _{q}n.  \label{def of q-kaijyo}
\end{equation}
\end{definition}

Using the definition of the $q$-product (\ref{def of q-product}), $\ln
_{q}\left( n!_{q}\right) \,$\ is explicitly expressed by $\ln _{q}\left(
n!_{q}\right) =\frac{\sum_{k=1}^{n}k^{1-q}-n}{1-q}.$If an approximation of $%
\ln _{q}\left( n!_{q}\right) $ is not needed, this explicit form should be
directly used for its computation. However, in order to clarify the
correspondence between the studies $q=1$ and $q\neq 1$, the approximation of 
$\ln _{q}\left( n!_{q}\right) $ is useful. In fact, using the following $q$%
-Stirling's formula, we obtain the unique generalized entropy corresponding
to the $q$-exponential function $\exp _{q}\left( x\right) $, shown in the
following sections.

\begin{theorem}
Let $n!_{q}$ be the $q$-factorial defined by (\ref{def of q-kaijyo}). The
rough $q$-Stirling's formula $\ln _{q}\left( n!_{q}\right) $ is computed as
follows:%
\begin{equation}
\ln _{q}\left( n!_{q}\right) =\left\{ 
\begin{array}{ll}
\dfrac{n}{2-q}\ln _{q}n-\dfrac{n}{2-q}+O\left( \ln _{q}n\right) \quad & 
\text{if}\quad q\neq 2, \\ 
n-\ln n+O\left( 1\right) & \text{if}\quad q=2.%
\end{array}%
\right.  \label{rough q-Stirling}
\end{equation}
\end{theorem}

The proof of the above formulas (\ref{rough q-Stirling}) is given in \cite%
{Su04b}.

\section{Tsallis entropy uniquely derived from the $q$-multinomial
coefficient and $q$-Stirling's formula}

In this section we show that Tsallis entropy is uniquely and naturally
derived from the fundamental formulations presented in the previous section.
In order to avoid separate discussions on the positivity of the argument in (%
\ref{def of q-multinomial coefficient}), we consider the $q$-logarithm of
the $q$-multinomial coefficient to be given by 
\begin{equation}
\ln _{q}\left[ 
\begin{array}{ccc}
& n &  \\ 
n_{1} & \cdots & n_{k}%
\end{array}%
\right] _{q}=\ln _{q}\left( n!_{q}\right) -\ln _{q}\left( n_{1}!_{q}\right)
\cdots -\ln _{q}\left( n_{k}!_{q}\right) .  \label{lnq of multinomial}
\end{equation}%
The unique generalized entropy corresponding to the $q$-exponential is
derived from the $q$-multinomial coefficient using the $q$-Stirling's
formula as follows \cite{Su04b}.

\begin{theorem}
When $n$ is sufficiently large, the $q$-logarithm of the $q$-multinomial
coefficient coincides with Tsallis entropy (\ref{d-Tsallis entropy}) as
follows:%
\begin{equation}
\ln _{q}\left[ 
\begin{array}{ccc}
& n &  \\ 
n_{1} & \cdots & n_{k}%
\end{array}%
\right] _{q}\simeq \left\{ 
\begin{array}{ll}
\dfrac{n^{2-q}}{2-q}\cdot S_{2-q}\left( \dfrac{n_{1}}{n},\cdots ,\dfrac{n_{k}%
}{n}\right) & \text{if}\quad q>0,\,\,q\neq 2 \\ 
-S_{1}\left( n\right) +\sum\limits_{i=1}^{k}S_{1}\left( n_{i}\right) & \text{%
if}\quad q=2%
\end{array}%
\right.  \label{important0-2}
\end{equation}%
where $S_{q}$ is Tsallis entropy \cite{Ts88}:%
\begin{equation}
S_{q}:=\frac{1-\sum\limits_{i=1}^{n}p_{i}^{q}}{q-1}
\label{d-Tsallis entropy}
\end{equation}%
and $S_{1}\left( n\right) $ is given by $S_{1}\left( n\right) :=\ln n.$
\end{theorem}

The proof of this theorem is given in \cite{Su04b}.

Note that the above relation (\ref{important0-2}) reveals a surprising 
\textit{symmetry}: (\ref{important0-2}) is equivalent to%
\begin{equation}
\ln _{1-\left( 1-q\right) }\left[ 
\begin{array}{ccc}
& n &  \\ 
n_{1} & \cdots & n_{k}%
\end{array}%
\right] _{1-\left( 1-q\right) }\simeq \frac{n^{1+\left( 1-q\right) }}{%
1+\left( 1-q\right) }\cdot S_{1+\left( 1-q\right) }\left( \frac{n_{1}}{n}%
,\cdots ,\frac{n_{k}}{n}\right)  \label{symmetry}
\end{equation}%
for $q>0$ and $q\neq 2$. This expression represents that there exists a 
\textit{symmetry} with a factor $1-q$ around $q=1$ in the algebra of the $q$%
-product. Substitution of some concrete values of $q$ into (\ref%
{important0-2}) or (\ref{symmetry}) helps us understand the meaning of this
symmetry.

Remark that the above correspondence (\ref{important0-2}) and the symmetry (%
\ref{symmetry}) reveals that the $q$-exponential function (\ref%
{q-exponential}) derived from\quad (\ref{nonlinear differential equation})
is consistent with Tsallis entropy only as information measure.

\section{The generalized Shannon additivity derived from the $q$-multinomial
coefficient}

This section shows another way to uniquely determine the generalized
entropy. More precisely, the identity derived from the $q$-multinomial
coefficient coincides with the generalized Shannon additivity which is the
most important axiom for Tsallis entropy.

Consider a partition of a given natural number $n$ into $k$ groups such as $%
n=\sum_{i=1}^{k}n_{i}$. In addition, each natural number $n_{i}\,\left(
i=1,\cdots ,k\right) $ is divided into $m_{i}$ groups such as $%
n_{i}=\sum_{j=1}^{m_{i}}n_{ij}$ where $n_{ij}\in \mathbb{N}$. 
\begin{figure}
[ptbh]
\begin{center}
\includegraphics[
height=1.93in,
width=3.69in
]%
{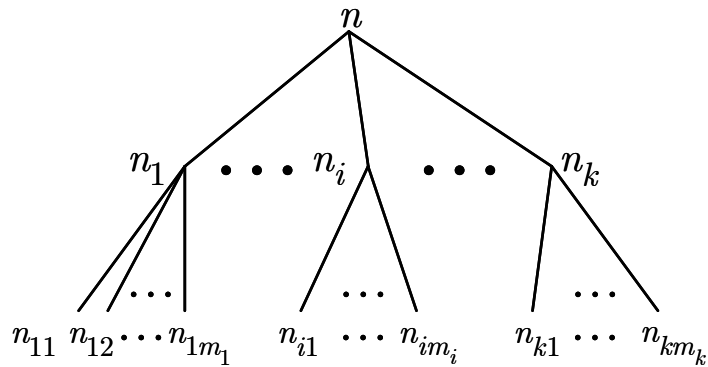}%
\caption{partition of a natural number $n$}%
\label{fig: partition}%
\end{center}
\end{figure}
Then, the following identity
holds for the $q$-multinomial coefficient.%
\begin{equation}
\quad \quad \left[ 
\begin{array}{ccc}
& n &  \\ 
n_{11} & \cdots & n_{km_{k}}%
\end{array}%
\right] _{q}=\left[ 
\begin{array}{ccc}
& n &  \\ 
n_{1} & \cdots & n_{k}%
\end{array}%
\right] _{q}\otimes _{q}\left[ 
\begin{array}{ccc}
& n_{1} &  \\ 
n_{11} & \cdots & n_{1m_{1}}%
\end{array}%
\right] _{q}\otimes _{q}\cdots \otimes _{q}\left[ 
\begin{array}{ccc}
& n_{k} &  \\ 
n_{k1} & \cdots & n_{km_{k}}%
\end{array}%
\right] _{q}  \label{q-identity}
\end{equation}%
It is very easy to prove the above relation (\ref{q-identity}) by taking the 
$q$-logarithm of the both sides and using (\ref{lnq of multinomial}).

On the other hand, the above identity (\ref{q-identity}) is reformed to the
generalized Shannon additivity in the following way. Taking the $q$%
-logarithm of the both sides of the above relation (\ref{q-identity}), we
have 
\begin{equation}
\ln _{q}\left[ 
\begin{array}{ccc}
& n &  \\ 
n_{11} & \cdots & n_{km_{k}}%
\end{array}%
\right] _{q}=\ln _{q}\left[ 
\begin{array}{ccc}
& n &  \\ 
n_{1} & \cdots & n_{k}%
\end{array}%
\right] _{q}+\sum_{i=1}^{k}\ln _{q}\left[ 
\begin{array}{ccc}
& n_{i} &  \\ 
n_{i1} & \cdots & n_{im_{i}}%
\end{array}%
\right] _{q}.
\end{equation}%
From the relation (\ref{important0-2}), we obtain%
\begin{equation}
S_{2-q}\left( \dfrac{n_{11}}{n},\cdots ,\dfrac{n_{km_{k}}}{n}\right)
=S_{2-q}\left( \dfrac{n_{1}}{n},\cdots ,\dfrac{n_{k}}{n}\right)
+\sum_{i=1}^{k}\left( \frac{n_{i}}{n}\right) ^{2-q}S_{2-q}\left( \dfrac{%
n_{i1}}{n_{i}},\cdots ,\dfrac{n_{im_{i}}}{n_{i}}\right) .
\label{n-gene-shanaddi}
\end{equation}%
Then, by means of the following probabilities defined by%
\begin{eqnarray}
p_{ij} &:&=\dfrac{n_{ij}}{n}\quad \left( i=1,\cdots ,k,\quad j=1,\cdots
,m_{k}\right) , \\
p_{i} &:&=\sum_{j=1}^{m_{i}}p_{ij}=\sum_{j=1}^{m_{i}}\dfrac{n_{ij}}{n}=\frac{%
n_{i}}{n}\quad \left( \because n_{i}=\sum_{j=1}^{m_{i}}n_{ij}\right) ,
\end{eqnarray}%
the identity (\ref{n-gene-shanaddi}) becomes%
\begin{equation}
S_{q}\left( p_{11},\cdots ,p_{km_{k}}\right) =S_{q}\left( p_{1},\cdots
,p_{k}\right) +\sum_{i=1}^{k}p_{i}^{q}S_{q}\left( \dfrac{p_{i1}}{p_{i}}%
,\cdots ,\dfrac{p_{im_{i}}}{p_{i}}\right) .  \label{generalizedShannon}
\end{equation}%
The formula (\ref{generalizedShannon}) obtained from the $q$-multinomial
coefficient is exactly same as the generalized Shannon additivity (See {%
[GSK3] given below}) which is the most important axiom for Tsallis entropy 
\cite{Su04d}.

In fact, the generalized Shannon-Khinchin axioms and the uniqueness theorem
for the nonextensive entropy are already given and rigorously proved in \cite%
{Su04d}. The present result (\ref{generalizedShannon}) and the already
established axiom {[GSK3]} perfectly coincide with each other.

\begin{theorem}
\bigskip Let $\Delta _{n}$ be defined by the $n$-dimensional simplex: 
\begin{equation}
\Delta _{n}:=\left\{ {\left( {p_{1},\ldots ,p_{n}}\right) \left\vert \;{%
p_{i}\geq 0,\;\sum\limits_{i=1}^{n}{p_{i}}=1}\right. }\right\} .
\label{simplex}
\end{equation}%
The following axioms [GSK1]$\sim $[GSK4] determine the function $%
S_{q}:\Delta _{n}\rightarrow \mathbb{R}^{+}$ such that 
\begin{equation}
S_{q}\left( {p_{1},\ldots ,p_{n}}\right) ={\frac{{{1-\sum\limits_{i=1}^{n}{%
p_{i}^{q}}}}}{{\phi \left( q\right) }}},  \label{theorem_Tsallis}
\end{equation}%
where $\phi \left( q\right) $ satisfies properties (i) $\sim $ (iv):
\end{theorem}

\begin{enumerate}
\item[(i)] $\phi \left( q\right) $ is continuous and has the same sign as $%
q-1,$i.e., 
\begin{equation}
\phi \left( q\right) \left( {q-1}\right) >0;  \label{conditionphi1}
\end{equation}

\item[(ii)] 
\begin{equation}
\underset{q\rightarrow 1}{\lim }\phi \left( q\right) =\phi \left( 1\right)
\!=\!0,\quad \phi \left( q\right) \neq 0\;\left( {q\!\neq \!1}\right) ;
\end{equation}

\item[(iii)] there exists an interval $\left( a,b\right) \subset \mathbb{R}%
^{+}$ such that $a<1<b$ and $\phi \left( q\right) $ is differentiable on the
interval 
\begin{equation}
\left( a,1\right) \cup \left( 1,b\right) ;
\end{equation}

\item[and] 

\item[(iv)] there exists a constant $k>0$ such that%
\begin{equation}
\underset{q\rightarrow 1}{\lim }\frac{{d\phi \left( q\right) }}{{dq}}=\frac{1%
}{k}.  \label{constrant4}
\end{equation}
\end{enumerate}

\begin{description}
\item {[GSK1]} \textit{continuity}: $S_{q}$ is continuous in $\Delta _{n}$
and $q\in \mathbb{R}^{+}$,

\item {[GSK2]} \textit{maximality}: for any $q\in \mathbb{R}^{+}$, any $n\in 
\mathbb{N}$ and any $\left( {p_{1},\cdots ,p_{n}}\right) \in \Delta _{n}$, 
\begin{equation}
S_{q}\left( {p_{1},\cdots ,p_{n}}\right) \leq S_{q}\left( {\frac{1}{n}}%
,\ldots ,{\frac{1}{n}}\right) ,
\end{equation}

\item {[GSK3]} \textit{generalized Shannon additivity}: if 
\begin{equation}
p_{ij}\geq 0,\;\;p_{i}=\sum\limits_{j=1}^{m_{i}}{p_{ij}}\;\;\forall
i=1,\cdots ,n,\forall j=1,\cdots ,m_{i},  \label{condition0}
\end{equation}%
then the following equality holds:%
\begin{equation}
S_{q}\left( p_{11},\cdots ,p_{nm_{k}}\right) =S_{q}\left( p_{1},\cdots
,p_{n}\right) +\sum_{i=1}^{n}p_{i}^{q}S_{q}\left( \dfrac{p_{i1}}{p_{i}}%
,\cdots ,\dfrac{p_{im_{i}}}{p_{i}}\right) ,  \label{q_Shannonadditivity}
\end{equation}
\end{description}

{[GSK4]} \textit{expandability}: 
\begin{equation}
S_{1}\left( {p_{1},\ldots ,p_{n},0}\right) =S_{1}\left( {p_{1},\ldots ,p_{n}}%
\right) .
\end{equation}

Note that, in order to uniquely determine the Tsallis entropy (\ref%
{d-Tsallis entropy}) in the above set of the axioms, \textquotedblleft $%
\underset{q\rightarrow 1}{\lim }$\textquotedblright\ should be removed from (%
\ref{constrant4}), that is, $\frac{{d\phi \left( q\right) }}{{dq}}=\frac{1}{k%
}$ (i.e., ${\phi \left( q\right) =}\frac{1}{k}\left( q-1\right) $) should be
used instead of (\ref{constrant4}). The general form ${\phi \left( q\right) }
$ perfectly corresponds to Tsallis' original introduction of the so-called
Tsallis entropy in 1988 \cite{Ts88}. See his original characterization shown
in page 9 of \cite{GT04} for the detail (${\phi \left( q\right) }$
corresponds to \textquotedblleft $a$\textquotedblright\ in his notation. His
simplest choice of \ \textquotedblleft $a$\textquotedblright\ coincides with
the simplest form of ${\phi \left( q\right) }$ i.e., $\frac{{d\phi \left(
q\right) }}{{dq}}=\frac{1}{k}.$).

When one of the authors (H.S.) submitted the paper \cite{Su04d} in 2002,
nobody presented the idea of the $q$-product. However, as shown above, the
identity on the $q$-multinomial coefficient \cite{Su04b} which was
formulated based on the $q$-product \cite{NMW03}\cite{Bo03} coincides with
one of the axioms (\textit{\ }{[GSK3]: generalized Shannon additivity) in }%
\cite{Su04d}. This means that the whole theory based on the $q$-product is
self-consistent. Moreover, other fundamental applications of the $q$%
-product, such as law of error \cite{Su04a} and the derivation of the unique
non self-referential $q$-canonical distribution \cite{WS05b}\cite{Su06}, are
also based on the $q$-product.

\section{Conclusion}

Starting from a fundamental nonlinear equation $dy/dx=y^{q}$, we present the
scaling property and the algebraic structure of its solution. Moreover, we
prove that the algebra determined by its solutions is mathematically
consistent with Tsallis entropy only as the corresponding unique information
measure based on the following 2 mathematical reasons: (1) derivation of
Tsallis entropy from the $q$-multinomial coefficient and $q$-Stirling's
formula, (2) coincidence of the identity derived from the $q$-multinomial
coefficient with the generalized Shannon additivity which is the most
important axiom for Tsallis entropy.

These mathematical discussions result in the self-consistency between the
mathematics derived from the $q$-exponential and Tsallis entropy.

Recently, we have presented the following fundamental results in Tsallis
statistics:

\begin{enumerate}
\item Axioms and the uniqueness theorem for the nonextensive entropy \cite%
{Su04d}

\item Law of error in Tsallis statistics \cite{Su04a}

\item $q$-Stirling's formula in Tsallis statistics \cite{Su04b}

\item $q$-multinomial coefficient in Tsallis statistics \cite{Su04b}

\item Central limit theorem in Tsallis statistics (numerical evidence only) 
\cite{Su04b}

\item $q$-Pascal's triangle in Tsallis statistics \cite{Su04b}

\item The unique non self-referential $q$-canonical distribution in Tsallis
statistics \cite{WS05b}\cite{Su06}

\item Scaling property characterized by the fundamental nonlinear
differential equation [the present paper].
\end{enumerate}

All of the above fundamental results are derived from the algebra of the $q$%
-product and mathematically consistent with each other. This means that the $%
q$-product is indispensable to the formalism in Tsallis statistics. More
important point is that the $q$-product originates from the fundamental
nonlinear differential equation (\ref{nonlinear differential equation}) with 
\textit{scale-invariant} solutions.


\begin{thebibliography}{99}
\bibitem{SW06} In any manuscripts and papers up to now, this constant $C$ is
set to be \textquotedblleft $0$\textquotedblright , i.e., $\exp _{q}\left(
C\right) =1$\ (See the equation (17) at page 5 of \cite{GT04} and\ the
equations (22)-(23) at page 8 of \cite{Ts04PD}). But the arbitrariness of
this constant $C$ plays a very important role in scaling (\ref{scaling}).
\end{thebibliography}
\end{document}